\documentclass[journal=nalefd,manuscript=letter]{achemso} %
\usepackage{graphicx}

\author{Thomas~Maassen}
	\email{t.maassen@rug.nl} 
\author{J.~Jasper~van~den~Berg}
\author{Natasja~IJbema}
  \affiliation[University of Groningen]
  {Physics of Nanodevices, Zernike Institute for Advanced Materials, University of Groningen, Nijenborgh 4, 9747 AG Groningen, The Netherlands}%
\author{Felix~Fromm}
\author{Thomas~Seyller}
	\affiliation[Friedrich-Alexander Universit\"{a}t Erlangen-N\"{u}rnberg]
	{Lehrstuhl für Technische Physik, Universität Erlangen-Nürnberg, Erwin-Rommel-Strasse 1, 91058 Erlangen, Germany}
\author{Rositza~Yakimova}
	\affiliation[Link\"{o}ping University]
	{Department of Physics, Chemistry and Biology (IFM), Link\"{o}ping University, S-581 83 Link\"{o}ping, Sweden}	
\author{Bart~J.~van~Wees}
  \affiliation[University of Groningen]
  {Physics of Nanodevices, Zernike Institute for Advanced Materials, University of Groningen, Nijenborgh 4, 9747 AG Groningen, The Netherlands}%

\date{2 December 2011}

\title[]
{Long spin relaxation times in wafer scale epitaxial graphene on SiC(0001).}
\begin{document}
\begin{abstract}
We developed an easy, upscalable process to prepare lateral spin-valve devices on epitaxially grown monolayer graphene on SiC(0001) and perform nonlocal spin transport measurements. We observe the longest spin relaxation times $\tau_S$ in monolayer graphene, while the spin diffusion coefficient $D_S$ is strongly reduced compared to typical results on exfoliated graphene. The increase of $\tau_S$ is probably related to the changed substrate, while the cause for the small value of $D_S$ remains an open question. 
\end{abstract}
\begin{figure}
\includegraphics[width=\columnwidth]{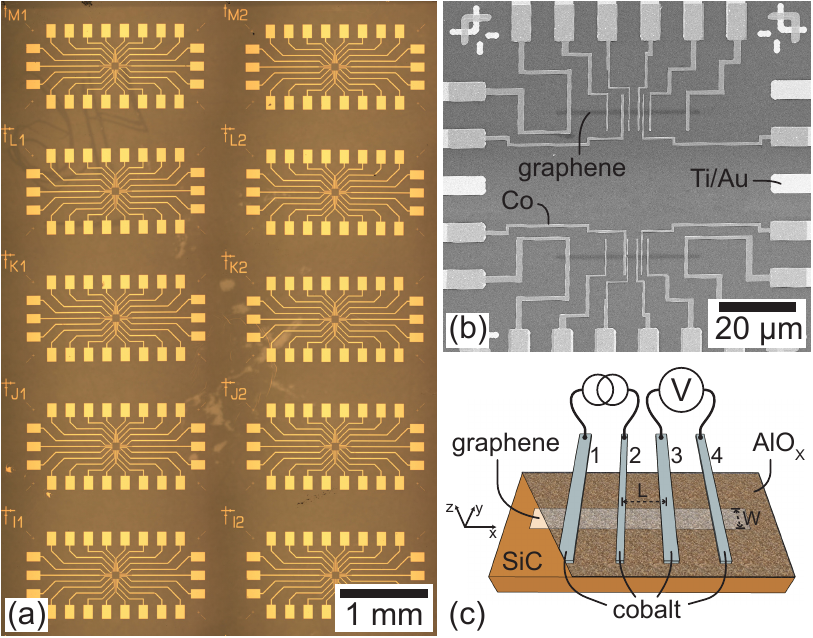} 
\caption{\label{fig:Fig1}(a) Optical microscope picture of a SiC wafer prepared with Ti/Au bondpads and leads to central $100 \times 100 ~\mathrm{\mu m^2}$ areas for further device preparation. Each of these patterns has a unique label for further production steps and measurements. The changes in the background color are due to scratches and resist residues on the backside of the transparent SiC wafer. (b) SEM picture of one of the central device areas with two spin valve devices, connected with Co electrodes to the Ti/Au leads. (c) Sketch of an MLEG spin valve device with four Co contacts. The wafer including the MLEG strip is covered with $\mathrm{AlO_x}$, before the Co contacts are deposited.}
\end{figure}
\begin{figure}
\includegraphics[width=\columnwidth]{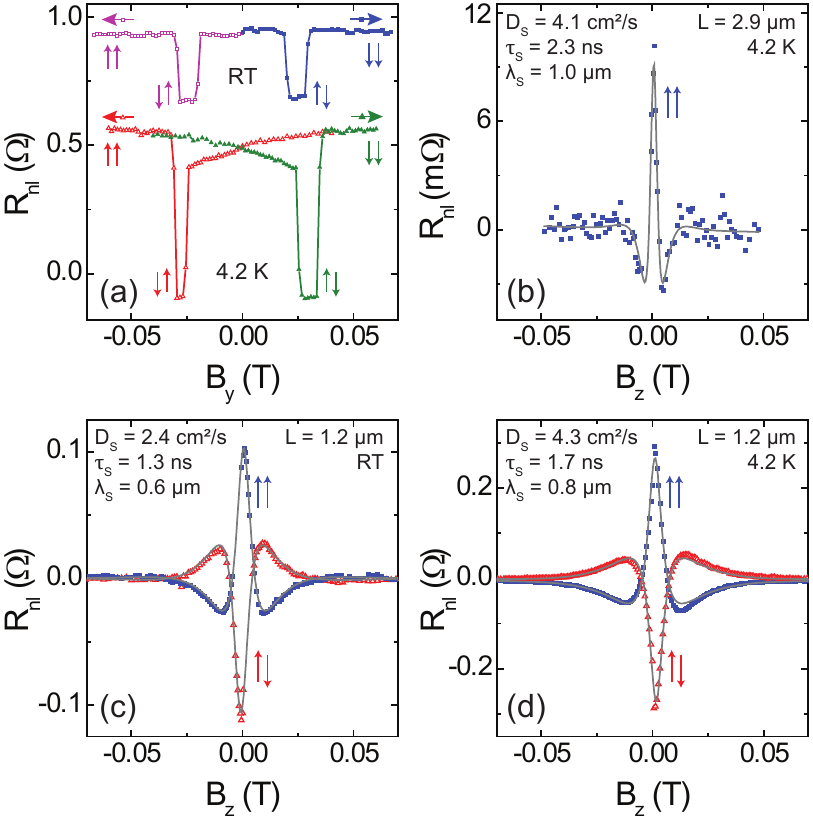} 
\caption{\label{fig:Fig2}Nonlocal spin transport measurements. (a) Spin valve measurements on a device with $L = 1.2 ~\mathrm{\mu m}$ inner contact distance at RT (purple and blue) and $4.2 ~\mathrm{K}$ (red and green). The sweep directions of the magnetic field are indicated by the horizontal arrows, the relative orientation of the Co contacts is illustrated by the pairs of vertical arrows. (b)-(d) Hanle precession measurement for parallel ($\uparrow \uparrow$, blue boxes) and antiparallel alignment ($\uparrow \downarrow$, red open triangles) of the inner electrodes for (b) $L=2.9 ~\mathrm{\mu m}$ at $4.2 ~\mathrm{K}$, (c) $L = 1.2 ~\mathrm{\mu m}$ at RT and (d) $L = 1.2 ~\mathrm{\mu m}$ at $4.2 ~\mathrm{K}$. The fits to the solutions of the Bloch equation are plotted in gray. The background resistance, which is visible in (a), is subtracted in the Hanle precession measurements (b)-(d).}
\end{figure}
\begin{figure}
\includegraphics[width=\columnwidth]{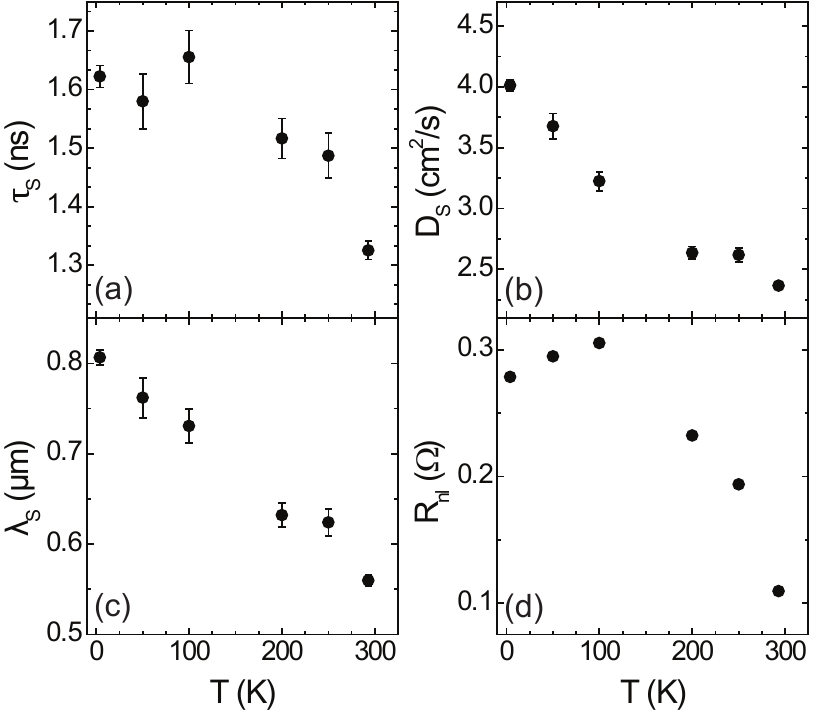} 
\caption{\label{fig:Fig3}Temperature dependence of (a) the spin relaxation time, (b) the spin diffusion coefficient, (c) the spin relaxation length and (d) the nonlocal signal for the sample with $L = 1.2 ~\mathrm{\mu m}$. If available, several fitting results at the same temperature were averaged.} 
\end{figure}

%\pacs{75.76.+j, 75.40.Gb}
%%%%%%%%%%%%%%%%%%%%%%%%%%%%%%%%%%%%%%%%%%%%%%%%%%%%%%%%%%%%%%%%%%%%%%%%%%%%%%%%%%%%%%%%%%%%%%%%%%%%%%%%%%%%%%%%%%%%%%%%%%%%%%%%%%%%%%%%%%%%%%%%%%%%%%%%%%%%%
%%%%%\section{\label{sec:Introduction}Introduction}
Spin transport in graphene draws great attention since the observation of spin relaxation lengths of $\lambda_S= 2~\mathrm{\mu m}$, with spin relaxation times in the order of $\tau_S = 150 ~\mathrm{ps}$ at room temperature (RT) in mechanical exfoliated single layer graphene (eSLG) \cite{N448_Tombros2007}. Recent experiments show an increase of $\tau_S$ to $\tau_S \approx 0.5 ~\mathrm{ns}$ in eSLG at RT \cite{PRL105_Han2010, PRL107_Han2011} and $\tau_S \approx 1 ~\mathrm{ns}$ at $T=4 ~\mathrm{K}$ \cite{PRL107_Han2011}. Measurements on bilayer graphene (BLG) show even higher spin relaxation times, up to a few nanoseconds at low temperature \cite{PRL107_Yang2011, PRL107_Han2011}. At the same time, a study on few-layer graphene (FLG) showed an enhancement of $\tau_S$ with increasing number of layers, which is attributed to the screening of external scattering potentials \cite{PRB83_Maassen2011}. \\
While most spin transport measurements were performed on exfoliated graphene, a first publication by Avsar et al.\cite{NL11_Avsar2011} showed measurements on graphene, grown by chemical vapor deposition (CVD) on copper foil. This publication marked the first step towards large scale production of spin transport devices, which showed similar spin transport properties compared to exfoliated graphene. The disadvantage of the growth of graphene on metal substrates is however that one is forced to transfer the material to an insulating substrate to be able to perform transport measurements. \\
Therefore it is useful to think about an alternative, e.g. epitaxially grown graphene on semi-insulating SiC \cite{MRSB35_First2010, PRB78_Virojanadara2008}. %
This letter is the first report of spin transport in this material and therefore the first report of spin transport in graphene on a different substrate than $\mathrm{SiO_2}$. We present lateral nonlocal spin-valve and spin-precession measurements on graphene strips prepared from monolayer epitaxial graphene (MLEG) grown on the Si-face of a semi-insulating SiC substrate (SiC(0001)) by sublimation of Si in Ar atmosphere \cite{PRB78_Virojanadara2008, NN5_Tzalenchuk2010, NM8_Emtsev2009}. \\
%
%
%%% \section{\label{sec:Sample_Fabrication}Sample Fabrication}
%
4H-SiC wafers \cite{ACSB25_Shaffer1969} are heated to $2000^\circ \mathrm{C}$ in an ambient argon pressure of 1 atm as described in Refs.~\citenum{PRB78_Virojanadara2008} and \citenum{NN5_Tzalenchuk2010}, leading to the growth of the so called buffer layer that is predominately ($>80\%$) covered with MLEG, with some areas uncovered or covered with double layer graphene. The measurements were performed on MLEG \cite{foot_MLEG} and with the help of Hall measurements on similar samples we estimate an electron doping with a charge carrier density of $n \approx 3 \times 10^{12} ~\mathrm{cm^{-2}}$ and a charge carrier mobility of $\mu \approx 1900 ~\mathrm{cm^2/Vs}$ at RT.\\
Figure~\ref{fig:Fig1}(a) shows an about $7\times5~\mathrm{mm^2}$ big part of a SiC wafer covered with MLEG, prepared with a pattern of Ti/Au structures that form a periodic pattern of bondpads with leads to central $100 \times 100 ~\mathrm{\mu m^2}$ areas for further device preparation. These structures are prepared in an optical lithography step, using a deep-UV mask aligner with a double resist layer (LOR-3A / ZEP-520A, from MicroChem / ZEON Corporation). After development, the wafer is etched with oxygen plasma at $40~\mathrm{W}$ for 20 seconds, before depositing a Ti/Au ($5~\mathrm{nm}/35~\mathrm{nm}$) double layer using e-beam evaporation followed by lift-off in PRS-3000 (from J.T. Baker). The etching step is necessary to enable the adhesion of the Ti/Au contacts on the substrate. To prepare the central device regions (Figure~\ref{fig:Fig1}(b)), two MLEG strips per area are defined, using e-beam lithography (EBL) on a negative resist (ma-N 2400, from micro resist technology GmbH) and the uncovered MLEG is etched in a second oxygen plasma etching step. % 
After this, the wafer is annealed for two hours in $\mathrm{Ar(95\%):H_2(5\%)}$ environment at $350^\circ\mathrm{C}$ to remove resist residues. To avoid the conductivity mismatch\cite{APS57_Fabian2007, PRB80_Popinciuc2009, unpublished_Maassen}, the wafer is covered with an approximately $1~\mathrm{nm}$ thick $\mathrm{AlO_x}$ layer by evaporating $0.4~\mathrm{nm}$ aluminum at a base pressure of $p < 1 \times 10^{-6}~\mathrm{mbar}$, oxidation in situ in $\mathrm{O_2}$ atmosphere at a base pressure of $p >3 \times 10^{-5}~\mathrm{mbar}$ for $15~\mathrm{min}$ and repeating the step a second time. Finally, in a standard PMMA resist based EBL step the $45~\mathrm{nm}$ thick Co electrodes are formed (Figure~\ref{fig:Fig1}(b)) connecting the Ti/Au leads with the two graphene strips, before the bondpads are contacted using wire bonding. By preparing the wafer with an optical step and using EBL only on the small central areas, we developed a fast and easy process to prepare a full wafer with (spin) transport devices. This process can also be used for different types of large area graphene on non-conducting substrates. \\
As the resolution of the ma-N resist is limited to about $50~\mathrm{nm}$, we developed a second process, replacing the ma-N resist step with a PMMA step. This enables a higher resolution but in return requires an additional step to remove the graphene outside the $100 \times 100 ~\mathrm{\mu m^2}$ areas to disconnect the Ti/Au leads. This method was implemented for two devices, using an additional optical lithography step, where we cover the central device areas, while the exposed graphene is removed with oxygen plasma. \\ 
%
%%%%%%%%\section{\label{sec:Spin_Transport}Spin Transport}
%
%
The presented measurements are performed on a $W = 0.7~\mathrm{\mu m}$ wide MLEG strip in vacuum at a base pressure of about $1 \times 10^{-6}~\mathrm{mbar}$ using low frequency lock-in technique and AC currents between $1$ and $10~\mathrm{\mu A}$. The measurements have been confirmed with consistent results, that have been obtained on several spin-valve areas on two other devices with $W = 1~\mathrm{\mu m}$. \\
The typical nonlocal geometry is presented in Figure~\ref{fig:Fig1}(c). A spin polarized current $I$ is sent from contact 2 to contact 1, generating a spin accumulation at contact 2, that diffuses in positive and negative x-direction. The $\mathrm{AlO_x}$ barrier separates the MLEG from the Co contacts and avoids reabsorption of the injected spins in the higher conducting cobalt \cite{PRB80_Popinciuc2009}. The exponential decaying spin accumulation generates a nonlocal voltage $V_{nl}$ between the spin sensitive contacts 3 and 4, which can be measured as a function of the magnetic field. In a spin-valve measurement, the magnetic field $B_y$, aligned with the contacts, is first used to bring the magnetization of the electrodes into a parallel (P) configuration and is then ramped in the opposite direction. When the magnetization of one of the electrodes is switched the measured voltage shows abrupt changes. The magnetic switching fields of the contacts are different due to different coercive fields which are achieved by different width of the contacts \cite{N448_Tombros2007, PRB80_Popinciuc2009}. \\
Figure~\ref{fig:Fig2}(a) shows two spin-valve measurements, one at RT and one at $4.2 ~\mathrm{K}$. $V_{nl}$, normalized to the nonlocal resistance $R_{nl}=V_{nl}/I$, is plotted as a function of the magnetic field. The upper measurement has been obtained at RT. After saturating the magnetization of the contacts at $B_x \approx -450 ~\mathrm{mT}$ no change of $R_{nl}$ is observed, before $B_y$ crosses $B_y=0$ (not shown). Then at $B_y=18~\mathrm{mT}$ a switch in $R_{nl}$ by $250~\mathrm{m \Omega}$ is observed, that can be attributed to the antiparallel alignment (AP) of injector and detector before it switches back to P and the initial $R_{nl}$ value at $B_y=30~\mathrm{mT}$. Based on only two visible switches it can be concluded that the outer contacts (contact 1 and 4 in Figure~\ref{fig:Fig1}(c)) are giving no significant contribution to the signal \cite{N448_Tombros2007}. The relatively small amplitude of the signal of $2 R_{nl} \approx 0.3~\mathrm{\Omega}$ is not necessarily related to spin relaxation in the graphene strip but is here related to the relatively low polarization of the contact interface. Also, the contact interface is described by the $R$ parameter, which is in our measurements $R=W~R_C/R_{sq}~\geq~2.1~\mathrm{\mu m}$, with a contact resistance of $R_C \geq 3.3~\mathrm{k\Omega}$, a square resistance of the MLEG of $R_{sq}=1.1~\mathrm{k\Omega}$ and $W=0.7~\mathrm{\mu m}$ \cite{PRB80_Popinciuc2009, foot_tempIndependent}. Therefore the contacts are almost noninvasive, but can still slightly influence the spin transport measurements \cite{PRB80_Popinciuc2009, unpublished_Maassen}. \\
The spin valve measurement, performed at $T=4.2 ~\mathrm{K}$, shows similar behavior. The main differences are, that the amplitude is approximately doubled and the switching fields are slightly increased, due to a change in the coercive fields with decreased temperature. Additional, $R_{nl}$ shows a gradual decrease in its value, before the contacts switch to AP. This is probably due to a slight misalignment of the magnetic field and the electrodes. The changed background resistance is mainly influenced by heat related effects \cite{c_Vera-Marun2011, PRL105_Bakker2010} and can therefore change with temperature. \\%
To analyze the spin transport properties, we perform Hanle spin precession measurements \cite{APS57_Fabian2007}. For this purpose the magnetic field is aligned in z-direction. The resulting spin dynamics are described with the one dimensional Bloch equation for the spin accumulation $\vec{\mu_S}$ \cite{APS57_Fabian2007}
\begin{equation}
D_S\mathbf\nabla^2 \vec{\mu_S} - \frac{\vec{\mu_S}}{\tau_S} + \vec{\omega_0}\times\vec{\mu_S}=\vec{0}.
\label{eq:Bloch}
\end{equation}
The first term on the left-hand side describes the spin-diffusion represented by the spin-diffusion coefficient $D_S$, and the second term describes the spin relaxation with the spin relaxation time $\tau_S$. The third term describes the precession with the Larmor frequency $\vec{\omega_0}=g\mu_B/\hbar \ \vec{B}$, where $g \approx 2$ is the effective Land\'{e} factor and $\mu_B$ is the Bohr magneton. \\
\begin{table}%
\centering
\begin{tabular}{|l|c|c|c|}
  \hline
  $L (\mathrm{\mu m})$ & $2.9$ & $1.2$ & $1.2$ \\
  \hline
  $T (\mathrm{K})$ & $4.2$ & $4.2$ & $293$  \\
  \hline
  $D_S (\mathrm{cm^2/s})$ & $4.06 \pm 0.05$ & $4.26 \pm 0.06$ & $2.38 \pm 0.03$\\
  \hline
  $\tau_S (\mathrm{ns})$ & $2.34 \pm 0.28$ & $1.66 \pm 0.02$ & $1.34 \pm 0.02$\\
  \hline
  $\lambda_S (\mathrm{\mu m})$ & $0.98 \pm 0.06$ & $0.84 \pm 0.01$ & $0.56 \pm 0.01$\\
  \hline
  $R_{nl} (\mathrm{m\Omega})$ & $9.2 \pm 0.4$ & $275.7 \pm 2.2$ & $102.9 \pm 0.7$\\
  \hline
\end{tabular}
\caption{Results of the fits to the measurements in Figure~\ref{fig:Fig2}(b)-(d).}
\label{tab:HanlePrecession}
\end{table}%
The Hanle precession measurements in Figure~\ref{fig:Fig2}(b)-(d) can be fitted with the solutions of the Bloch equation~(\ref{eq:Bloch}), yielding the spin transport quantities $D_S$ and $\tau_S$. A summary of the fitting results are shown in Table~\ref{tab:HanlePrecession}. Figure~\ref{fig:Fig2}(b) shows Hanle precession measurements, performed on a distance of $L=2.9 ~\mathrm{\mu m}$ at $4.2 ~\mathrm{K}$. The fit gives $D_S = 4.06 \pm 0.05~\mathrm{cm^2/s}$ and $\tau_S=2.34 \pm 0.28~\mathrm{ns}$, resulting in $\lambda_S=\sqrt{D_S \tau_S}= 0.98 \pm 0.06~\mathrm{\mu m}$. \\%
We would like to note, that this value for $\tau_S$ is the longest reported spin relaxation time on monolayer graphene. And as the contacts are to some extent invasive, we can expect even higher values for $\tau_S$, because a part of the injected spins relax at the contact interface. The effect of the contacts becomes apparent if one compares the fits of the measurement at $L=2.9 ~\mathrm{\mu m}$ and $L=1.2 ~\mathrm{\mu m}$ at $4.2~\mathrm{K}$ (Figure~\ref{fig:Fig2}(b) and (d)). For the measurement at $L=1.2~\mathrm{\mu m}$ we get $\tau_S=1.66 \pm 0.02~\mathrm{ns}$, which is around $70\%$ of the $\tau_S$ obtained from the $L=2.9 ~\mathrm{\mu m}$ measurement. This is due to the fact that the contact induced relaxation is more predominant, the shorter the distance the spins diffuse in the graphene strip between the contacts \cite{unpublished_Maassen}. With invasive contacts, the shorter measurement distance also leads to a slight increase in the measured $D_S$ \cite{unpublished_Maassen}. Also this is observed, comparing the $L=1.2~\mathrm{\mu m}$ to the $L=2.9 ~\mathrm{\mu m}$ precession at $4.2~\mathrm{K}$.\\%
When measuring at RT (Figure~\ref{fig:Fig2}(c)) we observe a reduction of $D_S$ by more than $40\%$ and $\tau_S$ is decreased by about $20 \%$. Therefore we get $\lambda_S= 0.56 \pm 0.01~\mathrm{\mu m}$, which is one third smaller than $\lambda_S$ at $4.2~\mathrm{K}$ (see Table~\ref{tab:HanlePrecession}). We also observe slightly higher values for $\tau_S$ of up to $\sim 1.5~\mathrm{ns}$ (not shown). Figure~\ref{fig:Fig3} shows a more detailed temperature dependence of $\tau_S$, $D_S$, $\lambda_S$ and the non-local signal amplitude $R_{nl}$ between $4.2~\mathrm{K}$ and RT. All four parameters show a decline between $4.2~\mathrm{K}$ and RT. While $D_S$ and $\lambda_S$ are monotonically decreasing by $40$ and $30 \%$, respectively, the value of $\tau_S$ and $R_{nl}$ drops by $20 \%$ and a factor of $3$, respectively. The decrease of all four values can be related to electron-phonon scattering \cite{APS57_Fabian2007, PRB84_Tanabe2011}. $\tau_S$ and $R_{nl}$ are approximately constant below $100~\mathrm{K}$ which could be related to the fact, that the phonons are frozen out below that temperature. Given the relatively low mobility of the graphene, the temperature dependence could also be dominated by Coulomb scattering on trapped charges in the SiC substrate which shows a strong temperature dependence as described by Farmer et al.~\cite{PRB84_Farmer2011}.\\
Our typical values for eSLG on $\mathrm{SiO_2}$ at RT are in case of $D_S$ about a factor of 80 bigger, however the measurements on MLEG strips show an increase of $\tau_S$ by about a factor of 10. This still leads to a $\sim70\%$ lower value for $\lambda_S$ \cite{PRB80_Popinciuc2009, PRB80_Jozsa2009}. The increase in $\tau_S$ in MLEG compared to eSLG can be attributed to the changed substrate. While $\mathrm{SiO_2}$ has an electrical inhomogeneous surface potential leading to electron-hole puddles \cite{NP4_Martin2008} and limiting effects for spin transport in graphene \cite{PRB80_Ertler2009}, the SiC crystal and the buffer layer are far more homogeneous and reduce therefore scattering. We would like to note that in our measurements on exfoliated graphene the spin transport properties are only weakly influenced by the temperature \cite{N448_Tombros2007, PRB83_Maassen2011} whereas we here see an improvement at low temperatures.\\
Although $\tau_S$ is improved, we do not know the origin for the reduced values of $D_S$. We obtain the diffusion coefficient $D_S$ from Hanle spin precession measurements. To verify if the value for $D_S$ is correct, we compare it to the diffusion coefficient $D_C$ acquired from charge transport measurements on the same area. $D_C$ is calculated using the Einstein relation $D_C=(R_{sq} e^2 \nu(E_F))^{-1}$, where $e$ is the electron charge and $\nu$ is the density of states (DOS). The band structure for MLEG on SiC(0001) is the same as for eSLG \cite{MRSB35_First2010}. Therefore, we can assume the same DOS as for eSLG, $\nu(E)=g_v g_s 2 \pi \left|E\right|/(h^2 v_F^2)$, with the twofold valley ($g_v=2$) and spin ($g_s=2$) degeneracies and the Fermi velocity $v_F\approx10^6~\mathrm{ms^{-1}}$. With $n$ estimated by Hall measurements on similar devices and $n(E_F)=\int^{E_F}_{0}\nu(E)dE$ we can calculate the Fermi energy $E_F$ and receive $\nu(E_F)$. With $n\approx3 \times 10^{12} ~\mathrm{cm^{-2}}$ and $R_{sq}=1.1~\mathrm{k\Omega}$ we get $D_C\approx190~\mathrm{cm^2/s}$ which is similar to typical values obtained in eSLG \cite{PRB80_Jozsa2009}. This is not surprising, because the charge carrier mobility in our samples is with $\mu=(R_{sq} e n)^{-1}\approx1900~\mathrm{cm^2/Vs}$ reasonably close to the mobility of our eSLG devices \cite{PRB80_Jozsa2009}. \\
%
%difference D_C and D_S
%
But this value of $D_C$ means that we observe a difference between the charge diffusion coefficient $D_C$ and $D_S$ of a factor 45 to 80 (compare Table~\ref{tab:HanlePrecession}) in contrast to $D_C \approx D_S$ in eSLG \cite{PRB80_Jozsa2009}. $D_S$ and $D_C$ can in principle be different as observed by Weber et al.\cite{N437_Weber2005} in a two-dimensional electron gas. Here the effect is attributed to electron-electron interactions but was significantly smaller than in our results. In FLG a slight difference of $\sim20\%$ between the two coefficients is found, \cite{PRB83_Maassen2011} but that difference is not comparable to the observation here. \\
We therefore do not expect a difference between the diffusion coefficient obtained from charge transport measurements and from Hanle precession measurements. While we cannot explain the observed difference, yet, we can exclude some possible explanations for it. \\
We do not expect the $D_C$ value to be incorrect as the observed charge transport is comparable to eSLG. One aspect though, that could result in a wrong $D_C$ value, are extra current paths next to the MLEG strip which would result in a change of the observed $R_{sq}$. We can exclude this by carefully controlling the etched structures with a scanning electron microscope (SEM) and by confirming that contacts of different MLEG strips show no conduction between each other.\\
$n$ was determined by Hall measurements on similar samples, but not on the spin transport samples themselves, therefore there could be an error in the value for $n$ which would lead to an incorrect value for the DOS. The highest values for $n$, measured on MLEG samples on SiC(0001), under the described growth conditions, are around $n \approx 1 \times 10^{13} ~\mathrm{cm^{-2}}$ which leads to $D_C\approx100~\mathrm{cm^2/s}$. This changes $D_C$ by less than a factor of $2$ and smaller values for $n$ would only increase the calculated value for $D_C$. Hence, also this aspect does not explain the difference between $D_S$ and $D_C$.\\
Another possibility would be a wrongly assumed DOS. Though, to result in values for $D_C$ similar to $D_S$, we would need a DOS as high as $\sim 50$ times the DOS of BLG. But we can be sure that we do not have such a DOS in our system because similar material to that used in our studies shows the typical quantum Hall effect (QHE) of eSLG \cite{NN5_Tzalenchuk2010, PRB81_Jobst2010}. \\
Since we do not find any explanation for the difference between $D_C$ and $D_S$ in the way $D_C$ is determined, let us have a look at $D_S$.
$D_S$ is obtained by the fit of the Hanle precession data in the same way as in earlier experiments \cite{N448_Tombros2007, PRB80_Jozsa2009, PRB80_Popinciuc2009, PRB83_Maassen2011}. Therefore the fitting procedure cannot explain the difference in the values, as the result for eSLG, $D_C \approx D_S$, is based on fits to measurements on eSLG using the same procedure. Next to that, we can confirm the value for $D_S$ in a different way. As mentioned before, the small value for $D_S$ leads to a relatively small value for $\lambda_S$. The order of magnitude of this value can be confirmed by comparing the change of $R_{nl}$ with $L$. By fitting an exponential decay \cite{APS57_Fabian2007} to the two $R_{nl}$ values versus $L$ for the data obtained at $4.2~\mathrm{K}$ (see Table~\ref{tab:HanlePrecession}, fit not shown), we receive $\lambda_S\approx0.5~\mathrm{\mu m}$ in agreement with the order of magnitude of $\lambda_S$ obtained from our fitted $\tau_S$ and $D_S$ \cite{foot_polarization}. With $D_S=D_C\approx 200~\mathrm{cm^2/s}$ and $\tau_S\approx 2~\mathrm{ns}$ we would receive a $\lambda_S$ of one order of magnitude larger. \\
While $\lambda_S$ is confirmed, there is still the possibility that the prefactor of the precession term in the Bloch equation~(\ref{eq:Bloch}) is wrong, which would affect linearly the determination of $D_S$ and $1/\tau_S$. This would be the case, if the effective Land\'{e} factor $g$ is changed in our system. But the reduction of $g$ by a factor of $\sim 50$ is needed, to result in our measured $D_S$. This is unlikely, also since $g\approx2$ was confirmed for epitaxial multilayer graphene on the C-face of SiC \cite{N467_Song2010}. \\
A change of $D_S$ can only be caused by the substrate as we expect the graphene to be comparable to eSLG \cite{MRSB35_First2010} and growth related defects like grain boundaries do not show a strong effect on $D_S$ and $\tau_S$ for CVD grown single layer graphene on $\mathrm{SiO}_2$ \cite{NL11_Avsar2011}. One of the substrate related effects could be inhomogeneities of the graphene thickness and doping at terrace step edges \cite{MRSB35_First2010} and scattering potentials resulting from that. However, this is relatively unlikely since step edges are not resulting in a discontinuity of the graphene layer \cite{MRSB35_First2010}. On the other hand, the out of plane electric field between the bulk SiC and the buffer layer \cite{MRSB35_First2010} could have an effect on the spin transport.\\
Another possible reason for the change in the spin transport properties could be related to the buffer layer. Its topology is graphene-like, though a part of the C atoms is covalently bonded to the underlying Si atoms. Therefore, the layer is electrically inactive and only weakly interacts with graphene layers on top \cite{PRB77_Emtsev2008}. The buffer layer does not seem to affect charge transport, at least not the measured resistance or the QHE \cite{MRSB35_First2010}, although it influences the temperature dependence of the charge carrier mobility \cite{APL99_Speck2011}. However, localized states in the buffer layer could act as hopping sites for electron spins and could influence the spin relaxation and the spin diffusion. By spins hopping into these states and back, $D_S$ could be reduced without affecting $R_{sq}$ and therefore the determined $D_C$ (as we do not include these extra states in the DOS). This kind of localized states could also originate from Al clusters in the $\mathrm{AlO_x}$ barrier. When depositing the barrier, some of the Al atoms could cluster and if their size exceeds some certain limit, part of the Al could stay non-oxidized. Those clusters would have a high DOS compared to the MLEG and could therefore have a relatively strong influence on the diffusion. This effect is unlikely, as we do not see it for eSLG on $\mathrm{SiO_2}$ but the less rough surface of MLEG on SiC and the resulting difference in the growth mechanism could result in this clustering. Here it would be interesting to produce samples with the $\mathrm{AlO_x}$ barrier only locally below the contacts as discussed in Ref.~\citenum{PRB80_Popinciuc2009} to compare the spin transport properties with the here reported results. \\
One other effect that could affect the measurements is the influence the $\mathrm{Ar(95\%):H_2(5\%)}$ cleaning at $350^\circ\mathrm{C}$ could have on the buffer layer. F. Speck et al.\cite{APL99_Speck2011} discuss the intercalation of hydrogen in epitaxial graphene on SiC which leads to the transformation of the buffer layer into an extra graphene layer. Though the discussed experiment uses about $1~\mathrm{bar}$ pure hydrogen at $550^\circ\mathrm{C}$ for 75 minutes, our cleaning step could partly intercalate hydrogen below the graphene layer and this could lead to extra transport channels and influence the transport measurements.\\
Non of these considerations above led to a conclusive explanation of the observed difference between the diffusion coefficients obtained from charge and spin transport measurements. Therefore we have to wait for further measurements to determine if the difference is based on the way those values are obtained or if there is a difference between charge and spin diffusion in MLEG on SiC(0001). The effect of the buffer layer can be addressed by measuring spin transport on quasi freestanding MLEG on SiC \cite{APL99_Speck2011} and general substrate related effects can be examined by transferring MLEG to $\mathrm{SiO_2}$ \cite{NL8_Lee2008}.\\ 
%
%%%%%%%%\section{\label{sec:Conclusions}Conclusions}
%
In summary we present a fast and easy process to prepare (spin) transport devices on wafer scale graphene by the example of MLEG. With this technique we produced lateral spin-valve devices on MLEG and performed spin-valve and Hanle spin precession measurements between $T=4.2~\mathrm{K}$ and RT. In the Hanle measurements we observe exceptionally high values for $\tau_S$ of up to $\tau_S=2.3~\mathrm{ns}$ and very small values for $D_S$ of $D_S < 5~\mathrm{cm^2/s}$, resulting in a reduction of $\lambda_S$ by a factor of $2$ to $3$ compared to eSLG. We observe a significant difference between the diffusion coefficient obtained from charge and spin transport measurements, which we discuss but cannot explain, yet. Finally we present the temperature dependence of the spin transport and show a decrease for $\tau_S$, $D_S$, $\lambda_S$ and $R_{nl}$ with increasing temperature, that can be linked to electron-phonon scattering or Coulomb scattering on trapped charges in the SiC substrate.\\

%%%% \begin{acknowledgments}
We would like to acknowledge P.~J.~Zomer, H.~M.~de~Roosz, B.~Wolfs, and J.~G.~Holstein for technical support and Sergey Kubatkin for helpful discussions. The research leading to these results has received funding from NanoNed, the Zernike Institute for Advanced Materials, the Foundation for Fundamental Research on Matter (FOM), the Deutsche Forschungsgemeinschaft and the European Union Seventh Framework Programme (FP7/2007-2013) under grant agreement ``ConceptGraphene'' Number 257829.
%%%% \end{acknowledgments}
%
\providecommand*\mcitethebibliography{\thebibliography}
\csname @ifundefined\endcsname{endmcitethebibliography}
  {\let\endmcitethebibliography\endthebibliography}{}


\begin{mcitethebibliography}{31}
\bibitem[foo()]{foot_MLEG}
The confirmation of the thickness to be MLEG is achieved by measuring
  $R_{sq}$.\relax
\mciteBstWouldAddEndPunctfalse
\mciteSetBstMidEndSepPunct{\mcitedefaultmidpunct}
{}{\mcitedefaultseppunct}\relax
\EndOfBibitem
\bibitem[foo()]{foot_tempIndependent}
Both $R_C$ and $R_{sq}$ show only weak temperature dependence, therefore also
  $R$ is not affected by $T$.\relax
\mciteBstWouldAddEndPunctfalse
\mciteSetBstMidEndSepPunct{\mcitedefaultmidpunct}
{}{\mcitedefaultseppunct}\relax
\EndOfBibitem
\bibitem[foo()]{foot_polarization}
As the polarization of the contact, and therefore the induced spin
  accumulation, can vary significantly between different contacts, this
  analysis is not conclusive, but allows a rough estimate of the order of
  magnitude of $\lambda_S$. The same analysis has been performed for the other
  two devices and has led to the same result.\relax
\mciteBstWouldAddEndPunctfalse
\mciteSetBstMidEndSepPunct{\mcitedefaultmidpunct}
{}{\mcitedefaultseppunct}\relax
\EndOfBibitem
\end{mcitethebibliography}
\end{document}